\documentstyle[pre,aps,multicol]{revtex}
\begin{document}

\draft
\title{About Probability Tails in Forced Burgers Turbulence}
\author{I. V. Kolokolov$^{+,*}$ and V. V. Lebedev$^{+,\#}$}
\address{$^+$ Physics of Complex Systems,
Weizmann Institute of Science, Rehovot 76100, Israel; \\
$^*$ Budker Institute of Nuclear Physics, 
Novosibirsk 630090, Russia; \\
$^\#$  Landau Inst. for Theor. Physics, Moscow, 
Kosygina 2, 117940, Russia.}
\date{20 December 2000}  

\maketitle

\begin{abstract}

We consider asymptotics of the velocity derivatives probability
distribution functions (PDFs) in Burgers' turbulence. We argue that
in the forced case the same power laws as in the decaying case are
realized for an infinite system.

\end{abstract}


\begin{multicols}{2}

The Burgers equation describes an evolution of weakly non-linear
one-dimensional acoustic waves in the reference frame moving with the
sound velocity \cite{Burg}. The equation is written for the velocity $u$
\begin{eqnarray} &&
\partial_t u+u\partial_x u
-\nu \partial_x^2 u=f \,,
\label{co1} \end{eqnarray}   
where $\nu$ is the viscosity coefficient and $f$ represents an external 
influence pumping the waves. The most interesting feature of this
equation is that it describes shock formation.

Recently statistical properties of solutions of the Burgers equation
(\ref{co1}) have attracted a considerable attention, since it is a simple
model which could demonstrate some features of developed turbulence
\cite{Frisch}. Then one assumes a large value of the Reynolds number 
${\rm Re}=VL/\nu$, where $V$ and $L$ are characteristic velocity and 
length of the large-scale motion. One can treat decaying ($f=0$) 
or forced turbulence. In the decaying case statistics of $u$ is
time-dependent, whereas it is expected to be stationary, if the pumping
force $f$ is a process statistically homogeneous in space and time.
We consider both possibilities.

An object, mainly treated in papers devoted to Burgers turbulence, is 
the PDF ${\cal P}_1(\xi)$ of the velocity derivative $\xi=u'$. This PDF
is strongly asymmetric. The right tail of the PDF decays fast in 
accordance with the inviscid law $\ln{\cal P}_1\propto-\xi^3$
\cite{Feigel,95Pol,96GM}. The inviscid left tail entailed a 
controversy. Different values of the exponent $\alpha$ for this 
tail ${\cal P}_1\propto|\xi|^{-\alpha}$ were proposed in the works
\cite{95Pol,96YC,96BM,97Bol,97EKMS,98GK,00BF}, see also the 
discussion in \cite{99EE,99Kra}. Below we confirm the validity of 
the answer $\alpha=7/2$, claimed in \cite{97EKMS,00BF}, for the forced
case in an infinite system.

We assume that the system size ${\cal L}$ is much larger than $L$. Then
the PDF of the velocity derivative $\xi_n=u^{(n)}$ can be written as the
space average
\begin{eqnarray} && 
{\cal P}_n(t,\xi_n)={\cal L}^{-1}
\int {\rm d}x\, \delta[\xi_n-u^{(n)}(t,x)] \,.
\label{pdf} \end{eqnarray}
The expression is the starting point of our consideration.
It is argued in the works \cite{97EKMS,00BF,00BFK} that the inviscid left 
tail of ${\cal P}_1(\xi)$ is related to the pre-shock regions. 
These vicinities possess some universal properties enabling to establish
the PDF (\ref{pdf}). 

We begin with the decaying case when $f=0$ in Eq. (\ref{co1}),
assuming a smooth initial condition $u_0(x)$ at $t=0$. This case
was examined by Bec and Frisch \cite{00BF}. Below, we reproduce their
arguments in a slightly different language.

It is well known that Lagrangian trajectories coming to the
shock creation points start at $t=0$ from points with zero second
derivative $u''_0=0$. First terms of the $u_0$ expansion near such 
point $X$ are
\begin{eqnarray} &&    
u_0\approx v+\sigma(x-X)
+\frac{s}{6}(x-X)^3 \,. 
\label{ex1} \end{eqnarray}
The shock is produced provided $\sigma<0$ and $s>0$. Clearly, the
contribution to the integral (\ref{pdf}) related to the shock is
independent of $X$ and $v$. Therefore the integral (\ref{pdf}) is
rewritten as
\begin{eqnarray} && 
{\cal P}_n(t,\xi_n)=\frac{\cal N}{\cal L}
\int{\rm d}\sigma\,{\rm d}s\, 
{\cal P}(\sigma,s)
\int {\rm d}x\, \delta[\xi_n-u^{(n)}] \,,
\label{pdr} \end{eqnarray}  
where ${\cal P}(\sigma,s)$ is the PDF of the parameters
$\sigma,s$ in the points $X_i$ and ${\cal N}$ is their number. The 
integral over $x$ in Eq. (\ref{pdr}) is taken over a vicinity of the
shock formation point related to the initial velocity profile (\ref{ex1}).
To be precise, a number of Lagrangian trajectories, starting from the
points $X_i$, are ``eaten'' by shocks, already formed, before coming to
the shock creation point. Such points $X_i$ have to be excluded from the
consideration, that leads to a $t$-dependence of both ${\cal N}$ and
${\cal P}(\sigma,s)$ in Eq. (\ref{pdr}).

The shock is formed at $t=|\sigma|^{-1}$. If $|\sigma|t<1$ then we can use
the characteristics method, which enables one to write $u(t,x)$ in the
parametric form
\begin{eqnarray} && 
x=y+ut \,, \qquad u=-|\sigma|y+\frac{s}{6}y^3 \,.
\nonumber \end{eqnarray} 
One can explicitly take the integral over $x$ in Eq. (\ref{pdr}) 
passing to integration over $y$. For large $\xi$ corresponding to
$1-|\sigma|t\ll1$ we get
\begin{eqnarray} &&  
\int {\rm d}x\, \delta[\xi-u'] 
=\sqrt{\frac{2}{s}}\,\frac{1}{|\xi|^3t^{5/2}}
\left[\frac{1}{t|\xi|}-(1-|\sigma|t)\right]^{-1/2} \,,
\nonumber \end{eqnarray} 
with the restriction $1-|\sigma|t<(t|\xi|)^{-1}$ (otherwise the
integral is zero). Then the integration over $\sigma$ leads to
\begin{eqnarray} &&   
{\cal P}_1(\xi)\sim\frac{\cal N}{\cal L}\,
\frac{1}{t^4|\xi|^{7/2}}
\int\frac{{\rm d}s}{\sqrt s}
{\cal P}(-t^{-1},s) \,.
\label{inv} \end{eqnarray} 
Thus we reproduce the answer obtained in the works \cite{97EKMS,00BF}.  
Analogously, the inviscid Khanin law $|\xi_n|^{-(3n+4)/(3n-1)}$ (written
in \cite{00BF}) can be obtained.

Now we proceed to investigating the forced Burgers turbulence.
We consider the force $f$ continuously distributed over time
and space and having a homogeneous statistics. Then the statistics of
$\xi_n$ is also homogeneous in time. We assume that $f$ is smooth 
both in space and time.          
 
As previously, we are interested in vicinities of the shock
formation points. Let us consider a prehistory of the $u$
evolution during the time $t$ before the shock creation point,
where $L/V\gg t\gg|\xi|^{-1}$. The first inequality enables one to
neglect the pumping force during the evolution. The second condition
ensures the validity of the expansion (\ref{ex1}) for examining the shock
formation. Thus, we return to the expression (\ref{inv}), where 
${\cal P}(\sigma,s)$ is the PDF of $\sigma=u'$ and $s=u'''$ in points
where $u''=0$, $u'<0$, $u'''>0$, this PDF is stationary in the pumping
case. However, there is a problem related to an explicit dependence of the
answer on the auxiliary time $t$. Let us demonstrate, that such dependence
is really absent.

The inequalities imposed on the evolution time $t$ mean that the values
$1\ll \sigma L/V,\ s L^3/V$ are relevant in the integral (\ref{inv}). We
assume also $\sigma L/V,\ s L^3/V \ll {\rm Re}^{1/2}$. In this range one
can neglect both pumping and the viscous term in Eq. (\ref{co1}). Then we
get from Eq. (\ref{co1}) a closed system of dynamical equations for
$\sigma=u'$ and $s=u'''$ (at the condition $u''=0$), leading to the
following equation for their PDF 
\begin{eqnarray} && 
\partial_t{\cal P}=
\partial_\sigma\left(\sigma^2{\cal P}\right)
+\partial_s\left(4\sigma s{\cal P}\right) \,,
\label{cw5} \end{eqnarray} 
For the stationary case the time derivative in Eq. (\ref{cw5}) is zero
and we get
\begin{eqnarray} && 
{\cal P}=\sigma^{-2} s^{-1}{\cal F}(s/\sigma^4) \,.
\label{cw6} \end{eqnarray} 
Of course, the function ${\cal F}(z)$ in (\ref{cw6}) knows
about pumping. Substituting the expression (\ref{cw6})
into Eq. (\ref{inv}) we get 
\begin{eqnarray} &&
{\cal P}_1(\xi)\sim\frac{\cal N}{\cal L}\,
\frac{1}{|\xi|^{7/2}}
\int\frac{{\rm d}z}{z^{3/2}}{\cal F}(z) \,.
\label{inv1} \end{eqnarray} 
We observe that the auxiliary time $t$ drops from the expression.
Thus, we reproduce the law $|\xi|^{-7/2}$ for the pumping case.
The same procedure gives the law $|\xi_n|^{-(3n+4)/(3n-1)}$ 
for high derivatives.

We are grateful to G. Falkovich for useful discussions and
to U. Frisch for valuable remarks.
We acknowledge a partial financial support of ISF.
I. V. K. is supported by RFFI grant 00-02-17652.

\end{multicols}

\end{document}